\documentclass[%
 reprint,
 superscriptaddress,
 amsmath,amssymb,
 aps,
prab,
]{revtex4-2}

\usepackage{graphicx}
\usepackage{dcolumn}
\usepackage{bm}
\usepackage{upgreek}
\usepackage{multirow}
\usepackage{algorithm}
\usepackage{algpseudocode}

\begin{document}

\preprint{APS/123-QED}

\title{Beam Optics Ramping in Under-Constrained Lattice Design: Application to Electron-Ion Collider Hadron Storage Ring Cooling Section}

\author{Derong Xu}\email{dxu@bnl.gov}
 \affiliation{Brookhaven National Laboratory}

\date{\today}

\begin{abstract}
This paper presents the lattice design and optics ramping strategy for the cooling section of the Hadron Storage 
Ring (HSR) at the Electron-Ion Collider (EIC). The main challenge is that available tuning knobs exceed beam-optics 
constraints. Independently optimized injection and top-energy optics often yield disconnected solutions, 
making interpolation impossible. To address this, we propose two new methods. 
The first is a midpoint-penalty scheme that ensures ramping path continuity by penalizing constraint violations 
at intermediate state.
The second is a top-down approach that adapts 
high-energy optics to low energy, guided by an adaptive weighting scheme to balance injection and ramping 
constraints. The solutions meet all beam dynamics and hardware limits. 
The two methods offer a general strategy for ramping in systems where the solution space is under-constrained and 
the starting and target configurations are far apart.
\end{abstract}

\maketitle

\section{Introduction}\label{sec:introduction}

Beam optics ramping is a fundamental task in synchrotron accelerators. The lattice must evolve smoothly between 
operational states to maintain acceptable optics. 
Common examples include the beta squeeze in colliders \cite{SolfaroliCamillocci:IPAC2016-TUPMW031}, where the beta functions at the interaction point (IP) 
are reduced gradually, and solenoid ramping in detectors \cite{morozov:ipac2024-mopc43}, which requires 
continuous optics matching to minimize betatron coupling as the solenoid field increases. 

A particularly challenging case arises in the Electron-Ion Collider (EIC), where the cooling section of the 
Hadron Storage Ring (HSR) must support electron cooling requirements at injection and accommodate different 
conditions at top energy. The large variation in requirements complicates the construction of a smooth and 
feasible ramping path.

\begin{figure}[htbp]
    \centering
    \includegraphics[width=0.9\columnwidth]{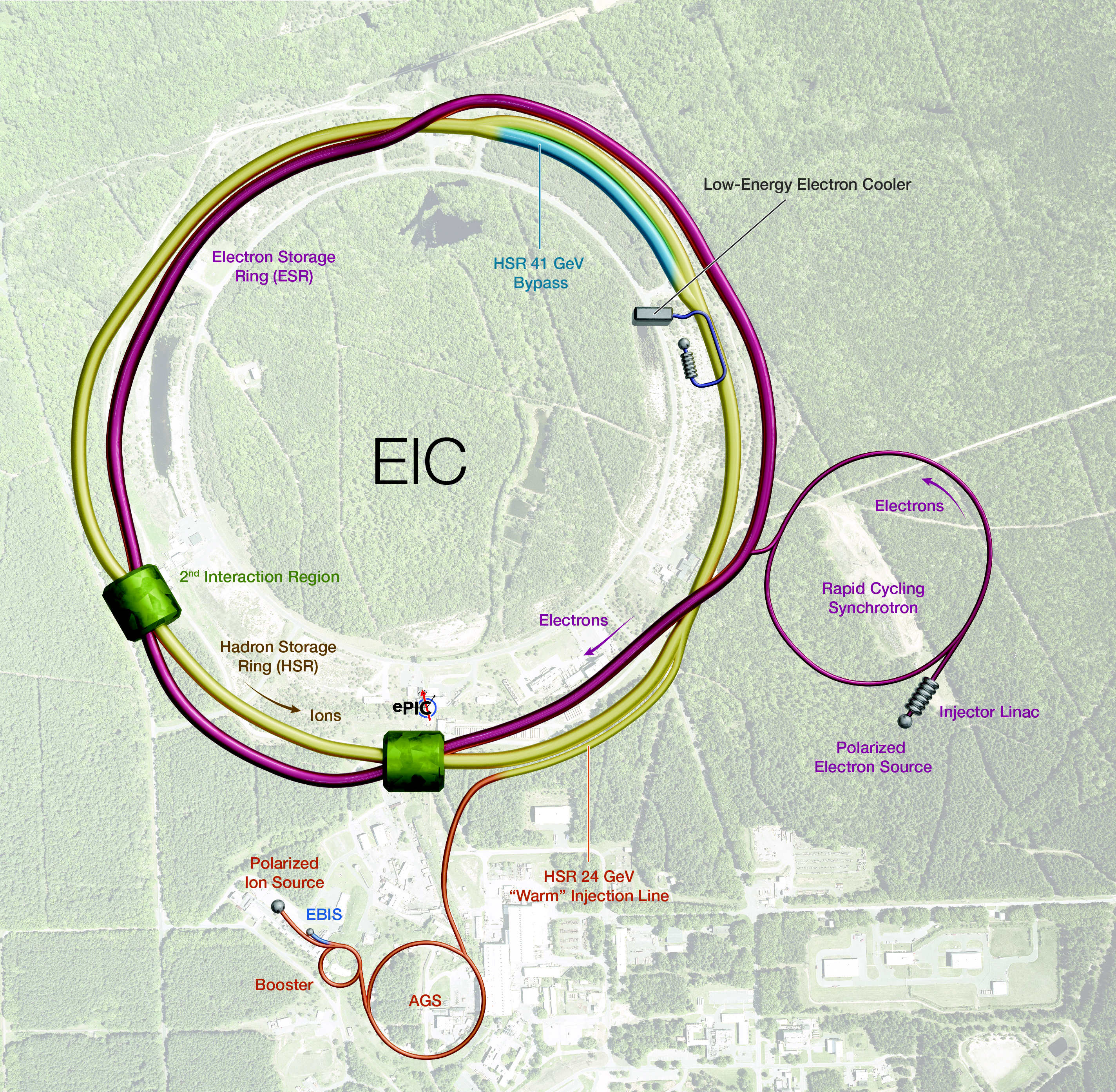}
    \caption{Conceptual layout of the EIC, showing key components including the HSR, ESR, injector complex, 
    and interaction regions. IR2 serves as the cooling section for generating flat hadron beams and 
    probably support high-energy cooling in future upgrades.}
    \label{fig:EIC-scheme}
\end{figure}

The EIC, under construction at Brookhaven National Laboratory, will collide polarized 
electron and hadron beams over a broad energy and species range. It targets a peak luminosity of 
$10^{34}~\mathrm{cm}^{-2}\mathrm{s}^{-1}$  and average polarization of $\sim 70\%$ \cite{osti_1765663}. 
As shown in Fig.~\ref{fig:EIC-scheme}, the EIC includes multiple key components: the Hadron Storage Ring (HSR), 
the Electron Storage Ring (ESR), a new injector complex, and two interaction regions.

The HSR, upgraded from Relativistic Heavy Ion Collider (RHIC), will store highly polarized 
hadron beams with a flat profile, where vertical emittance ten times smaller than horizontal. Six helical 
Siberian Snakes \cite{Mane_2005} will preserve polarization during acceleration and storage. 
The Insertion Region at two 
o’clock (IR2)\footnote{This paper follows the same naming convention for other regions, such as IR8.} will serve as 
the cooling section to generate flat beams at injection energy and potentially provide high-energy cooling. 
It will also host one Siberian Snake.

This paper is organized as follows. Section~\ref{sec:problem} defines the beam optics matching and ramping problem 
for HSR-IR2, including relevant constraints. 
Section~\ref{section:ramping:withoutRC} presents a midpoint-penalty method for the case without top-energy 
cooling constraints.
Section~\ref{section:ramping:withRC} addresses the more constrained case and 
develops a top-down strategy that adapts high-energy optics to satisfy injection requirements.
Section~\ref{sec:summary} summarizes the results.
\section{Problem}\label{sec:problem}
A flat hadron beam is essential for achieving the EIC’s peak luminosity goal. 
According to the EIC design, the injection and top-energy optics must satisfy distinct optics requirements, 
and a smooth ramping path is required to preserve the emittance ratio during acceleration.
\subsection{EIC cooling strategy}
Synchrotron radiation is negligible in the EIC HSR. Without external cooling, beam emittance is determined by the 
source and is similar in both transverse planes. The original EIC strategy included staged cooling: a Low energy 
Electron Cooler (LEC) is used at injection to create a flat hadron beam \cite{fedotov2020experimental}, followed 
by strong hadron cooling (SHC) at top energy. Two methods were considered for the SHC --- Micro-Bunched 
Electron Cooling (MBEC) \cite{bergan2021design} and electron cooling via a Ring Cooler (RC)
\cite{PhysRevAccelBeams.24.043501}. However, luminosity studies show that the SHC increases integrated luminosity by 
only a factor of two \cite{bergan:ipac2024-thyd1}. The current EIC baseline therefore defers high-energy cooling 
while preserving layout flexibility for a future upgrade.

Once the flat hadron beam is established at injection, it must be accelerated to top energy through a ramping process. 
Although a large transverse emittance ratio has been demonstrated at RHIC \cite{PhysRevLett.132.205001}, preserving 
beam flatness during ramping has not been experimentally verified. This motivates careful optics design to suppress 
vertical emittance growth. Accurate control of orbit, coupling, and Twiss functions is essential. In general, the 
ramping should be adiabatic, with continuously matched optics to minimize emittance dilution. The inclusion of the SHC
strongly impacts ramping design. In the absence of high-energy cooling, there are no specific optics constraints at top 
energy, allowing greater flexibility. In contrast, if high-energy cooling is implemented, the IR2 optics must satisfy 
strict conditions set by the chosen cooling method, complicating the construction of a smooth, matched ramping path. 
This paper uses the Ring Cooler as an example to compare ramping strategies with and without 
top-energy cooling constraints.

\subsection{Constraints and formulation}
The HSR-IR2 layout must satisfy geometric constraints. A primary goal is to allocate a 
long drift for low-energy electron cooling while accommodating a Siberian Snake and remaining compatible with the 
RHIC tunnel and magnet inventory. The layout should reuse existing RHIC superconducting magnets to reduce cost and 
simplify integration. It also respects the modularity of RHIC magnet families, while allowing flexibility through
the module rearrangement. 

\begin{figure*}[htbp]
    \centering
    \includegraphics[width=0.95\linewidth]{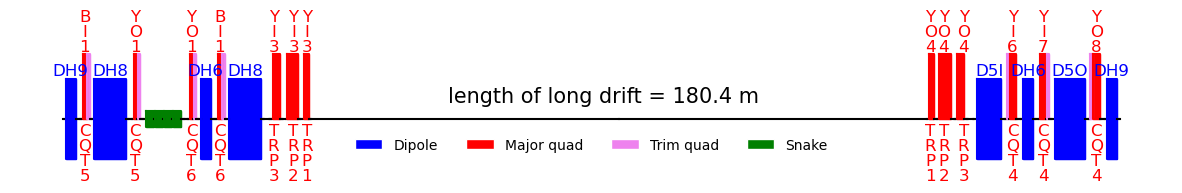}
    \caption{Optimized HSR-IR2 layout used for lattice design. This configuration achieves the longest possible usable 
    drift section. 
    The placement of dipoles, quadrupoles, and the Siberian 
    snake is shown schematically. Quadrupole and dipole names are labeled near their respective magnet blocks. 
    These names follow RHIC naming conventions; for further details, refer to the RHIC Configuration Manual~\cite{RHICConfigManual}.}
    \label{fig:HSR-IR2-layout}
\end{figure*}

The configuration shown in Fig. \ref{fig:HSR-IR2-layout} meets these requirements and serves as the geometric 
basis for the beam optics matching discussed in the following sections. Additional layout details are provided 
in Appendix \ref{app:layout}.

The beam optics must satisfy Twiss and dispersion matching conditions at arc 
interfaces, limits on beam size and aperture, and quadrupole strength bounds. 
These constraints are summarized in Appendix~\ref{app:constraints}.

Given these conditions, the optics matching problem at a given energy is formulated as solving a system of nonlinear equations 
under box and inequality constraints:
\begin{equation} 
    \mathbf{f}(\mathbf{k}) = \mathbf{0} 
    \quad \text{subject to} \quad \mathbf{k}_l \leq \mathbf{k} \leq \mathbf{k}_u, 
    \quad \mathbf{c}(\mathbf{k}) \leq \mathbf{0}, 
    \label{eq:problem}
\end{equation} 
where $\mathbf{k} \in \mathbb{R}^n$ is the vector of quadrupole strengths, 
$\mathbf{f}(\mathbf{k}) \in \mathbb{R}^m$ represents the exact matching conditions 
(e.g., Twiss and dispersion functions), $\mathbf{c}(\mathbf{k})$ encodes 
soft constraints such as aperture or beam size limits, and 
$\mathbf{k}_l, \mathbf{k}_u$ are the hardware-imposed lower and upper 
bounds on $\mathbf{k}$. 

In this paper, the equality constraints are normalized on a comparable scale:
\begin{equation}
    F_i=\frac{f_i-\overline{f}_i}{\mathrm{max}\left(\epsilon,\left|\overline{f}_i\right|\right)}
    \label{eq:normalizedObjective}
\end{equation}
where $f_i$ is the quantity to be matched,  
$\overline{f}_i$ stands for the target value, and $\epsilon$
is a user-defined floor to prevent over-weighting small targets; we use $\epsilon=0.01$.

Similarly, inequality constraints are handled using a one-sided penalty formulation. 
Specifically,
\begin{equation}
    F_{m+i} = \frac{\max\left(0,c_i - \overline{c}_i\right)}{\max\left(\epsilon,\left|\overline{c}_i\right|\right)},
    \label{eq:normalizedConstraint}
\end{equation}

The matching is performed by minimizing the objective vector $\mathbf{F}$ using the Levenberg–Marquardt algorithm \cite{LeastSquaresOptim}, 
or minimizing $||\mathbf{F}||$ via global search methods such as differential evolution \cite{BlackBoxOptim}.

The variable space $\mathbf{k}$ consists of $n=20$ tunable knobs. The number of hard matching conditions, i.e., 
independent components of $\mathbf{f}(\mathbf{k})$, is typically $m=9$ or $12$, 
making the system under-determined. This redundancy allows multiple valid solutions at fixed energies, 
offering flexibility for optics matching.

However, it complicates ramping. Solutions at injection and top energy may occupy disconnected volumes of $k$-space. 
In addition, the bounds $\mathbf{k}_l$  and $\mathbf{k}_u$ are 
energy-dependent; a solution feasible at one energy may fall outside the feasible range at next energy step. 
As a result, the optimization 
could fail to converge or yield discontinuous solutions, disrupting the continuity of the ramping path. 
These challenges motivate the need for a robust ramping strategy.

\section{Matching for LEC only}\label{section:ramping:withoutRC}

In this case, the beam optics at the center of the long drift is only constrained at 
injection. According to Table~\ref{tab:CombinedOptics}, the LEC requires the beta functions 
to lie within the range $100~\mathrm{m} \leq \beta_{x,y}^* \leq 200~\mathrm{m}$, 
where the superscript ``*'' denotes at the center of the long drift. 
However, the aperture constraints listed in Table~\ref{tab:ApertureConstraints} 
strongly favor solutions near the lower bound of this range. Consequently, 
the optimizer is naturally driven toward $\beta_{x,y}^* \approx 100~\mathrm{m}$.

The LEC also allows a range for the horizontal dispersion and its derivative, 
as shown in Table~\ref{tab:CombinedOptics}. However, due to the large beta functions 
in this region, the optics are effectively close to a dispersion-free condition. 
Therefore, the matching constraints are simplified by imposing:
\begin{equation}
    \beta^*_{x,y}=100~\mathrm{m},\quad \alpha_{x,y}^*=0,\quad
    \eta_x^*=0,\quad\eta_x'^*=0
    \label{eq:LEC-EqualityConstraints}
\end{equation}

\begin{figure}[htbp] 
    \centering 
    \includegraphics[width=0.99\linewidth]{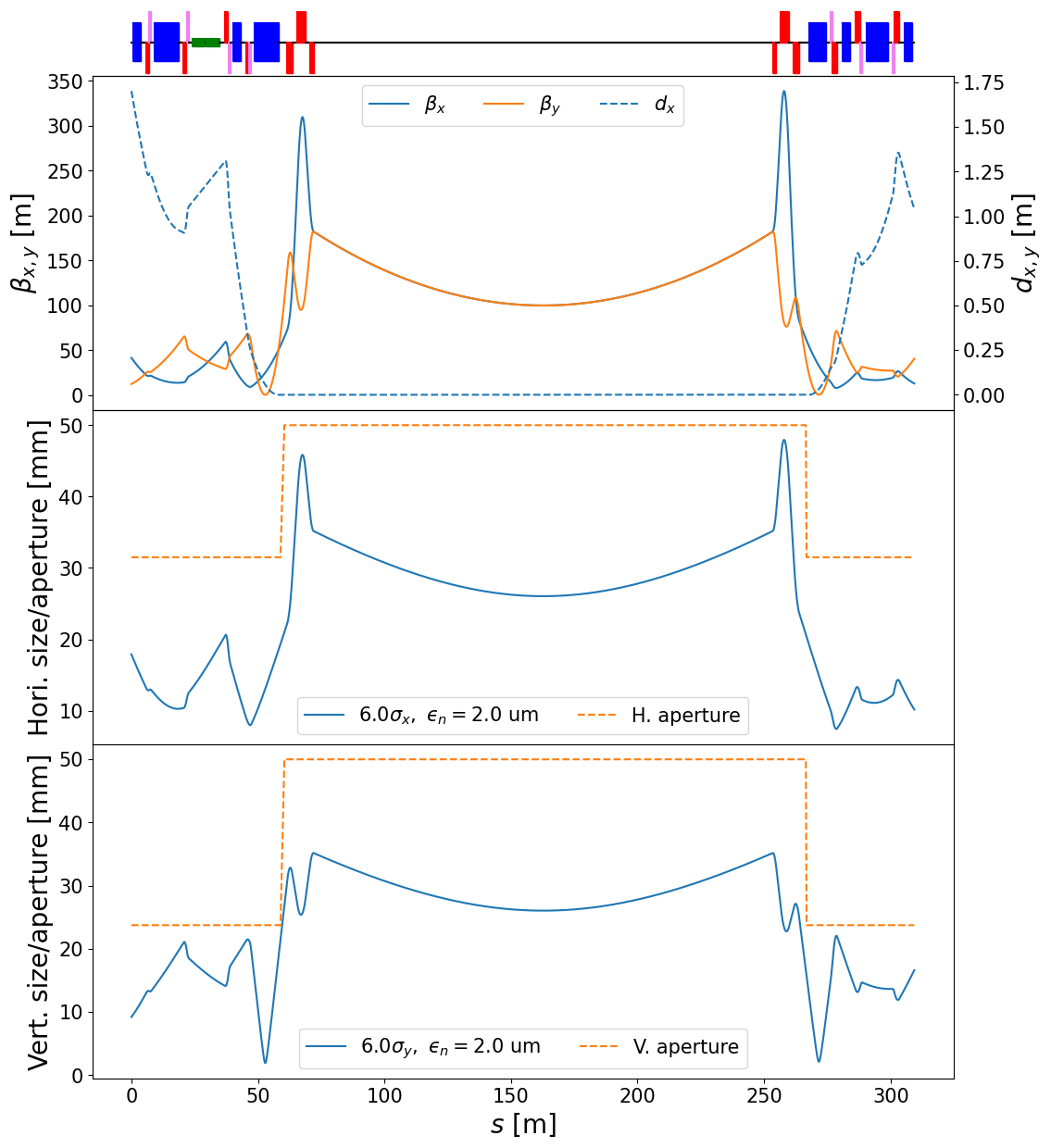} 
    \caption{Injection beam optics without RC constraints. Top: $\beta$-functions 
    and horizontal dispersion; Middle: horizontal beam size and aperture limit; 
    Bottom: vertical beam size and aperture limit.}
    \label{fig:FitLEC_withoutRC} 
\end{figure}

Figure \ref{fig:FitLEC_withoutRC} shows the matched optics solution along the 
IR2 region. The Twiss parameters $\beta_x$, $\beta_y$, $\eta_x$, and $\eta_x'$ 
at the center of the long drift, as well as at the region boundaries, satisfy 
the target values specified in Eq.(\ref{eq:LEC-EqualityConstraints}) and 
Table~\ref{tab:CombinedOptics}. The aperture requirements listed in 
Table~\ref{tab:ApertureConstraints} are also fulfilled throughout the region.

Although the beam optics can be successfully matched at injection energy, 
the resulting solution introduces challenges for constructing a smooth ramping path. 
In particular, many trim quadrupoles operate with opposite polarity to 
their neighboring main quadrupoles. This configuration effectively forms 
local doublets, where the trim and main quadrupoles act in opposition to 
control the horizontal and vertical beam sizes simultaneously.

However, these quadrupole pairs are positioned very close to each other, 
resulting in nearly identical phase advances. From the optics perspective, 
they function almost as a single element, and to achieve any significant 
differential effect, both quadrupoles must operate at high strength to 
produce a small net focusing. 

In the matched solution shown in Fig.~\ref{fig:FitLEC_withoutRC}, 
the strongest quadrupole pair operates at normalized focusing strengths 
of $0.19378~\mathrm{m}^{-2}$ and $-0.18541~\mathrm{m}^{-2}$, respectively. 
These values are several times larger than the hardware-imposed limits at top energy.
It is therefore impossible to ramp these quadrupoles directly with increasing energy.

Although the beam optics at top energy is not strictly constrained in this scenario,
it is beneficial to impose additional reasonable conditions to ensure the optics
well-behaved and avoids extreme configurations. 

As always, the Twiss parameters
at both boundaries listed in Table~\ref{tab:CombinedOptics} must be matched.
In addition, the following conditions are introduced:
\begin{equation}
    \alpha_{x,y}^*=0,\quad
    \eta_x'^*=0,\quad
    \beta_{x,y}<400~\mathrm{m}
    \label{eq:topEnergyOpticsWithoutRC}
\end{equation}
Here, the conditions $\alpha_{x,y}^* = 0$ and $\eta_x'^* = 0$ ensure symmetry of
the beta and dispersion functions across the long drift. The constraint
$\beta_{x,y} < 400~\mathrm{m}$ is consistent with the injection beam optics in 
Fig.~\ref{fig:FitLEC_withoutRC} and maintains a reasonable upper bound on the beta 
functions.

After obtaining the matched optics at top energy, a common strategy for 
constructing an energy ramp is to interpolate between the injection and 
top-energy solutions. This is the approach used in the RHIC ramping system, 
where Step Stones store set points for each device, and cubic interpolation is 
applied between stones to define intermediate values \cite{795328}. 

However, this strategy implicitly assumes that
the matched optics at both ends lie on a continuous solution branch in the
variable space. In our case, the constraints defined in Eq.~(\ref{eq:topEnergyOpticsWithoutRC}) do not guarantee that the injection 
and top-energy configurations are connected in $\mathbf{k}$-space. Therefore, 
direct interpolation between them often fails to produce a feasible ramping path.

\begin{figure}[htbp] 
    \centering 
    \includegraphics[width=0.99\linewidth]{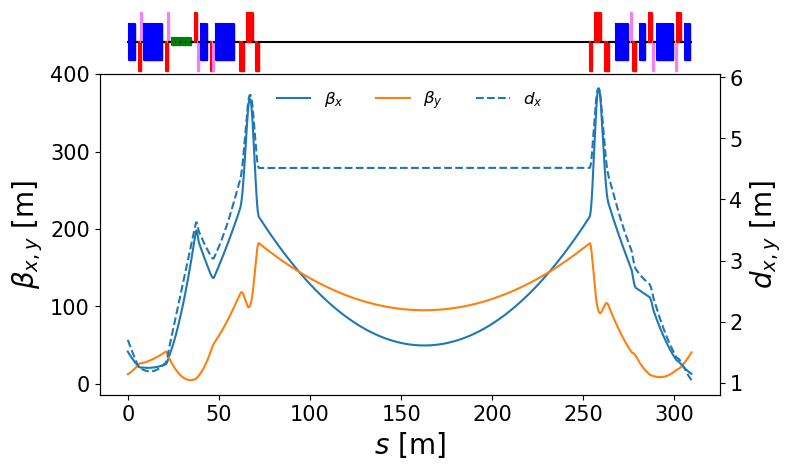} 
    \includegraphics[width=0.99\linewidth]{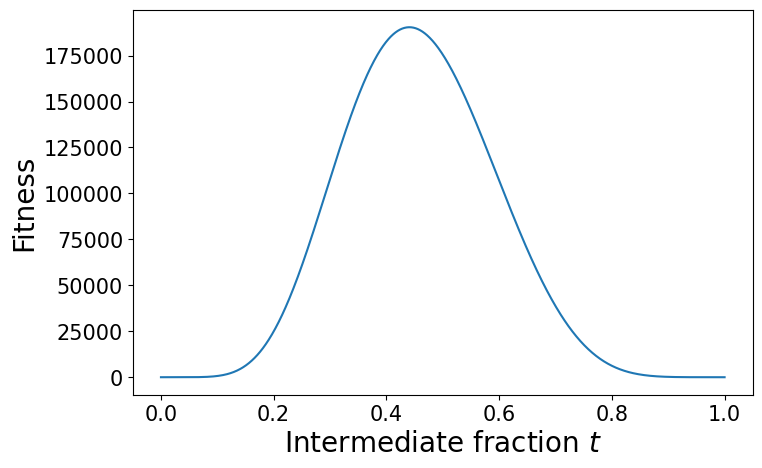} 
    \caption{Matched top-energy ($110~\mathrm{GeV}$) beam optics configuration 
    that cannot be connected to the injection solution through interpolation.  
    Top: Matched $\beta$-functions and horizontal dispersion at top energy, 
    satisfying the constraints in Eq.~(\ref{eq:topEnergyOpticsWithoutRC}) 
    and Table~\ref{tab:CombinedOptics}.  
    Bottom: Fitness evaluated at intermediate ramping fractions using 
    linear interpolation in quadrupole strength, as defined 
    in Eq.~(\ref{eq:linearFitQuadrupoleConfiguration}).  
    The large fitness in the middle of the ramp indicates that this top-energy 
    solution is not feasible.}
    \label{fig:withoutRC_infeasible} 
\end{figure}

Figure~\ref{fig:withoutRC_infeasible} presents an example where the top-energy optics 
is successfully matched, but the solution is not continuously connected to the 
injection optics obtained earlier. The top plot shows the Twiss parameters at 
top energy, satisfying both the matching conditions in Eq.(\ref{eq:topEnergyOpticsWithoutRC}) and 
the boundary conditions listed in
Table \ref{tab:CombinedOptics}.

In the bottom plot of Fig.~\ref{fig:withoutRC_infeasible}, we evaluate the fitness function at intermediate ramping fractions by linearly interpolating the quadrupole strengths between injection and top configurations. Specifically, the interpolated strength vector is defined as:
\begin{equation}
    \mathbf{k}(t)=
    t\cdot\mathbf{k}_\mathrm{top}
    +(1-t)\cdot
    \mathbf{k}_\mathrm{inj},
    \quad
    0\leq t\leq1
    \label{eq:linearFitQuadrupoleConfiguration}
\end{equation}
where $t=0$ corresponds to injection configuration $\mathbf{k}_\mathrm{inj}$ and 
$t=1$ to top configuration $\mathbf{k}_\mathrm{top}$
The fitness is evaluated as the squared 2-norm of the objective 
vector: $||\mathbf{F}||^2$.

When $t$ is close to $0$ or $1$, 
the fitness remains near zero, as the interpolated quadrupole configuration stays 
close to the respective matched solutions. However, the fitness increases to a large 
number in the middle between them. 
As shown in Eq.~(\ref{eq:normalizedObjective})–(\ref{eq:normalizedConstraint}), 
the objective and constraint terms have been normalized, 
yet the fitness at intermediate state exceeds $10^5$, indicating a severe 
violation of the matching conditions.

This large increase demonstrates that the interpolated quadrupole configuration 
passes through a region of parameter space where the optics constraints cannot 
be satisfied. Importantly, this behavior persists across different 
interpolation schemes, including smooth alternatives such as cubic splines.
It suggests that the injection 
and top-energy optics lie on disconnected branches of the feasible solution manifold 
in $\mathbf{k}$-space. 

Even if another valid ramping path exists, it would be 
located far from the region defined by the two endpoints and require entirely 
different quadrupole settings. Given the hardware-imposed range constraints, 
such a trajectory would be physically inaccessible. Therefore, this matched top-energy
configuration in Fig.~\ref{fig:withoutRC_infeasible} is infeasible.

To ensure that the injection and top-energy optics configurations can be connected 
through a feasible and continuous ramping path, we modify the objective function 
for optimizing the top-energy solution by adding a penalty term at an 
intermediate ramping fraction $t=0.5$. Specifically, the optimization is expressed as:
\begin{equation}
\begin{gathered}
    \mathbf{k}_\mathrm{mid}=\frac{\mathbf{k}_\mathrm{inj}+\mathbf{k}_\mathrm{top}}{2}\\
    \underset{\mathbf{k}_\mathrm{top}}{\text{minimize}}\ 
    \left\{||\mathbf{F}(\mathbf{k}_\mathrm{top})||^2
    + \mathrm{max}\left(0,\frac{||\mathbf{F}\left(\mathbf{k}_\mathrm{mid}\right)||-\delta}{\delta}\right)^2\right\}
\end{gathered}
\end{equation}
where $\delta$ is a user-defined tolerance; we set $\delta=m$ equal to the number of 
hard matching constraints defined in Eq.(\ref{eq:topEnergyOpticsWithoutRC}) 
and Table~\ref{tab:CombinedOptics}.

\begin{figure}[htbp] 
    \centering 
    \includegraphics[width=0.99\linewidth]{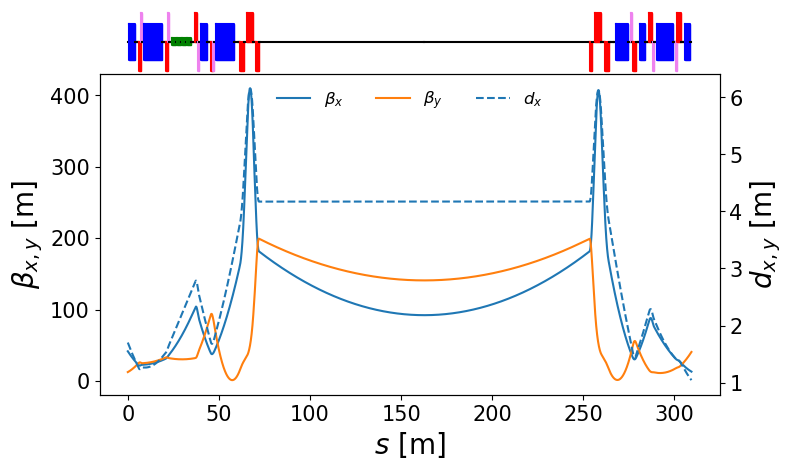} 
    \includegraphics[width=0.99\linewidth]{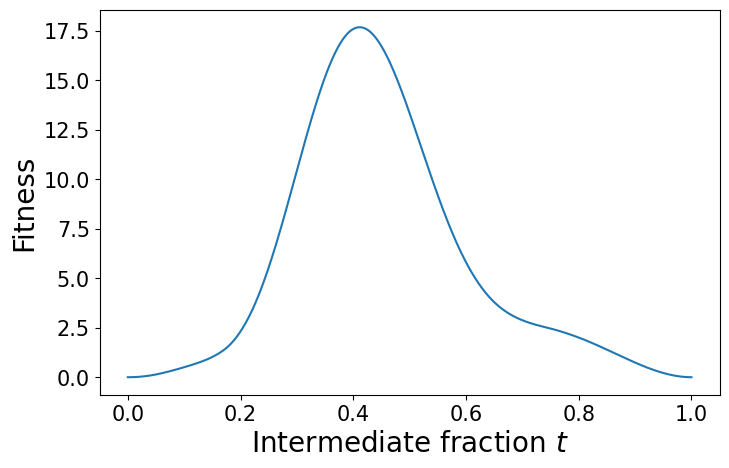} 
    \caption{Matched top-energy optics ($110~\mathrm{GeV}$) with feasible connection to injection solution.
    Top: Matched $\beta$-functions and horizontal dispersion at top energy, 
    satisfying the constraints in Eq.~(\ref{eq:topEnergyOpticsWithoutRC}) and 
    Table~\ref{tab:CombinedOptics}.  
    Bottom: Fitness $||\mathbf{F}\left((\mathbf{k}(t)\right)||^2$ as a function of intermediate 
    fractions during linear interpolation in 
    quadrupole strength, as defined by 
    Eq.~(\ref{eq:linearFitQuadrupoleConfiguration}).  
    The fitness remains low across the ramp, demonstrating a smooth and 
    feasible connection between injection and top-energy optics.}
    \label{fig:withoutRC_feasible} 
\end{figure}

The penalty becomes active only when the fitness at the midpoint exceeds the tolerance $\delta$.
It promotes smooth variation of $||\mathbf{F}(\mathbf{k}(t))||$ along the ramp.
A large midpoint violation implies a sharp change in constraint violation with respect to $t$, 
corresponding to a large Lipschitz constant of $||\mathbf{F}(\mathbf{k}(t))||$ \cite{bertsekas2009convex}, 
and suggests that the feasible region is disconnected along the interpolation path.

This approach effectively biases the optimizer toward configurations that support a continuous and physically realistic ramping trajectory.
Figure~\ref{fig:withoutRC_feasible} shows one such top-energy solution, where the interpolated quadrupole settings 
yield low fitness values at all intermediate steps, confirming the existence of a smooth and valid ramp.

\begin{figure}[htbp] 
    \centering 
    \includegraphics[width=0.99\linewidth]{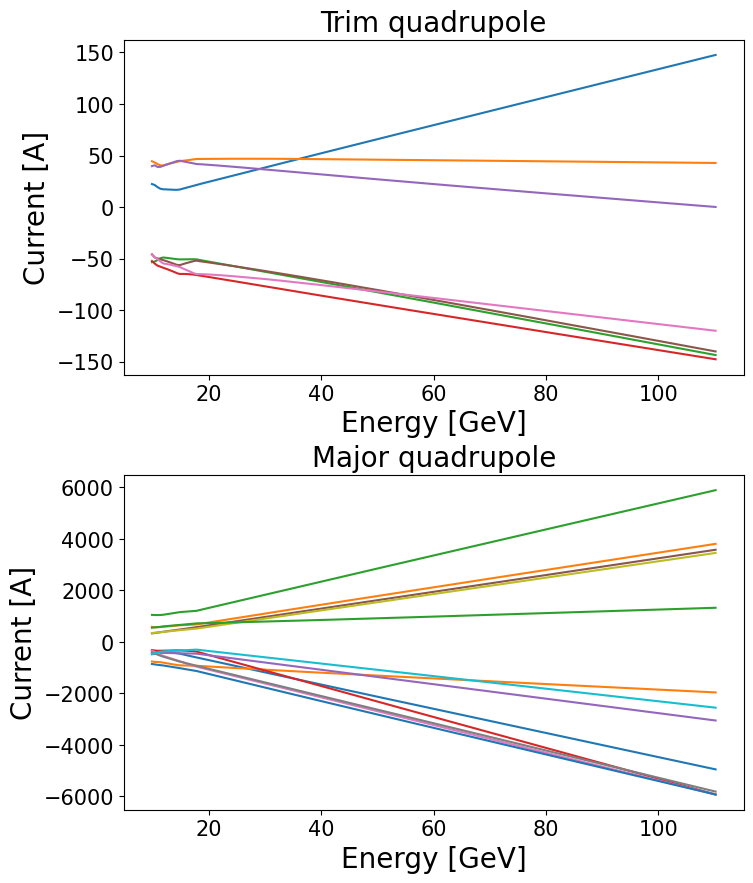} 
    \caption{Quadrupole currents during the energy ramping from injection to 
    top energy, top: the current evolution of the trim 
    quadrupoles, bottom: current of the main quadrupoles. 
    Each curve corresponds to an individual quadrupole in the IR2 region.}
    \label{fig:withoutRC_ramping} 
\end{figure}

Based on the feasible solution shown in Fig. \ref{fig:withoutRC_feasible}, a set of 
stones is selected to serve as anchor points in the ramping process. 
Between stones, a cubic spline interpolation is applied to the quadrupole current 
to ensure smooth variation within hardware limits. The beam energy is incremented 
in uniform steps of $0.5~\mathrm{GeV}$. At each energy step, a local optimizer 
is applied to fine-tune the quadrupole strengths so that the full set of 
optics constraints of Eq.~(\ref{eq:topEnergyOpticsWithoutRC}) 
and Table~\ref{tab:CombinedOptics} is satisfied. The final ramping current 
profiles for all quadrupoles are illustrated in Fig.~\ref{fig:withoutRC_ramping},
demonstrating that the proposed method is practically effective.
    
\section{Matching for LEC\&RC}\label{section:ramping:withRC}
In this scenario, the LEC and the RC impose different optics requirements at the center of the 
long drift, as shown in Table~\ref{tab:CombinedOptics}. As a result, the strategy described 
in Section~\ref{section:ramping:withoutRC}, which defines a common fitness constraint across the entire ramping process, 
is no longer applicable. 

If only loose constraints are imposed (e.g., $\beta_y < 1000~\mathrm{m}$), 
they fail to sufficiently reduce redundancy, offering little guidance to the optimizer. 
Conversely, enforcing strict constraints --- such as requiring that the Twiss and dispersion functions at the drift 
center interpolate smoothly between the injection and top-energy optics --- can over-constrain the problem and render 
it intractable, given the bounded quadrupole strength.

Most critically, without clearly defined intermediate constraints, it becomes impossible to determine whether the 
injection and top-energy optics lie on a connected solution branch in $\mathbf{k}$-space. 
The previous midpoint-penalty strategy cannot be directly applied in this setting.

\begin{figure}[htbp] 
    \centering 
    \includegraphics[width=0.99\linewidth]{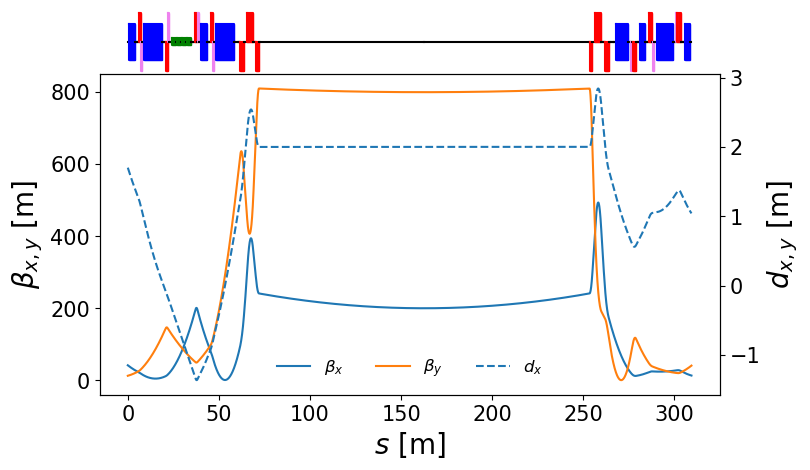} 
    \caption{Matched top-energy ($110~\mathrm{GeV}$) beam optics configuration 
    satisfying 
    quadrupole strength limitation in Table~\ref{tab:QuadStrengthLimits}.
    and
    the beam optics constraints in Table~\ref{tab:CombinedOptics}.}
    \label{fig:withRC_top} 
\end{figure}

Instead of independently matching the beam optics at injection and top energy and constructing the ramping path 
by interpolating the quadrupole strengths $\mathbf{k}$, we adopt a different strategy. 
To avoid disconnected solutions in $\mathbf{k}$-space, we begin by matching the top-energy optics, 
as shown in Fig.~\ref{fig:withRC_top}, and then gradually transition toward satisfying the LEC constraints at injection.

This top-down approach is motivated by two considerations. First, the top-energy requirements listed in 
Table~\ref{tab:CombinedOptics} are more difficult to satisfy. Second, the allowed quadrupole strength range is wider 
at lower energy, ensuring that the high-energy solution remains within bounds as the energy decreases. 
This helps prevent local optimization from failing due to initial guesses violating variable limits.

The injection constraints are categorized into two groups. The first, $\mathbf{F}_{\text{ramp}}$, includes the 
boundary conditions in Table~\ref{tab:CombinedOptics} and must be satisfied throughout the entire ramp. 
The second, $\mathbf{F}_{\text{inj}}$, includes the aperture constraints in Table~\ref{tab:ApertureConstraints} and 
the low-energy optics targets at the drift center, also defined in Table~\ref{tab:CombinedOptics}, and is enforced 
only at injection energy.

To implement the gradual transition, we introduce a weighting parameter $\lambda$ on the injection-specific 
constraints $\mathbf{F}_\mathrm{inj}$. Instead of minimizing the full objective vector $\mathbf{F}$, 
we minimize the augmented objective vector:
\begin{equation}
    \mathbf{F}_\lambda\left(\mathbf{k}\right)=\left[
    \begin{matrix}
        \mathbf{F}_\mathrm{ramp}\\
        \lambda\cdot\mathbf{F}_\mathrm{inj}
    \end{matrix}
    \right]
\end{equation}
where $\lambda$ is the tunable weight controlling the importance of injection-specific constraints in the optimization.

As shown in algrothm \ref{alg:adaptive_lambda}, an adaptive strategy is used to automatically select an 
appropriate $\lambda$ that balances feasibility and 
convergence. Starting from a small $\lambda$, we iteratively update it using a randomized scaling factor:
\begin{equation}
\lambda \gets \lambda \times 2^{\mathrm{rand()} - 0.49}
\end{equation}
where \texttt{rand()} generates a random number uniformly in $(0,1)$.
This introduces bidirectional perturbation, with a slight bias toward increasing $\lambda$.
A fixed $\lambda$ often causes the algorithm to converge to a local minimum. 
In contrast, the randomized scaling promotes exploration of the solution space and helps the optimizer escape local minima.

At each iteration, the Levenberg–Marquardt algorithm is applied within a small neighborhood around the current 
solution to update the quadrupole strengths $\mathbf{k}$. 

If the resulting fit exceeds a high threshold, $\lambda$ is halved, and the process repeats until
an acceptable $\lambda$ is found. If the fit falls below a low threshold, $\lambda$ is doubled to 
further enforce injection constraints. 

The iterative process terminates when $\lambda$ exceeds a preset upper 
limit (e.g., $10$), indicating that the injection optics are sufficiently well approximated while all 
ramping constraints are maintained.

\begin{algorithm}[H]
\caption{Adaptive $\lambda$-Weighted Optics Matching Strategy for Ramping Constraints}
\label{alg:adaptive_lambda}
\begin{algorithmic}[1]
\Require $\lambda_\mathrm{max}=10$
\Require Top-energy solution $\mathbf{k}_{\mathrm{top}}$, range bounds $\mathbf{k}_l, \mathbf{k}_u$
\Require Constraint function $\mathbf{F}_\lambda(\mathbf{k}) = \left[\mathbf{F}_{\mathrm{ramp}}^\top,\, \lambda \cdot \mathbf{F}_{\mathrm{inj}}^\top\right]^\top$
\Require Search radii of quadrupole strength in each step $\Delta \mathbf{k} = 10^{-3}~\mathrm{m}^{-2}$
\Require Tolerances $\texttt{lo\_tol}=10^{-3}$ (target fit quality) and $\texttt{hi\_tol}=10^{-2}$ (acceptable maximum during $\lambda$ search)
\State Initialize $\lambda \gets 1$, $\mathbf{k} \gets \mathbf{k}_{\mathrm{top}}$
\While{$\lambda < \lambda_{\text{max}}$}
    \State $\lambda \gets \lambda \times 2^{\mathrm{rand()} - 0.49}$ \Comment{Randomized $\lambda$ scaling to promote exploration}
    \State Compute local bounds: $\mathbf{k}_l' = \max(\mathbf{k} - \Delta \mathbf{k}, \mathbf{k}_l)$, $\mathbf{k}_u' = \min(\mathbf{k} + \Delta \mathbf{k}, \mathbf{k}_u)$
    \State Run local optimizer to solve $\min_{\mathbf{k}} ||\mathbf{F}_\lambda(\mathbf{k})||^2$ within bounds
    \State Record new solution $\mathbf{k}'$ and objective $J = ||\mathbf{F}_\lambda(\mathbf{k}')||^2$
    \While{$J > \texttt{hi\_tol}$}
        \State $\lambda \gets \lambda / 2$
        \State Re-optimize with updated $\lambda$
    \EndWhile
    \If{$J < \texttt{lo\_tol}$}
        \State $\lambda \gets 2 \cdot \lambda$ \Comment{Bias toward increasing $\lambda$ if fit is very good}
    \EndIf
    \State Update: $\mathbf{k} \gets \mathbf{k}'$
\EndWhile
\end{algorithmic}
\end{algorithm}

\begin{figure}[htbp] 
    \centering 
    \includegraphics[width=0.99\linewidth]{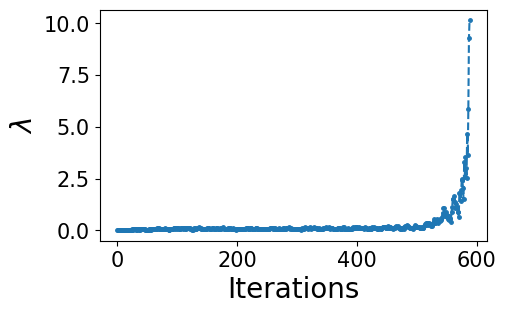} 
    \caption{Evolution of the weight parameter $\lambda$ during adaptive optimization. The stochastic update allows the optimizer to escape local minima and better balance between the ramping and injection constraints.}
    \label{fig:withRC_lambda} 
\end{figure}

The evolution of $\lambda$ during the optimization process is illustrated in Fig.~\ref{fig:withRC_lambda}. 
For the majority of iterations, $\lambda$ remains small and exhibits oscillatory behavior, allowing for 
sufficient exploration of the solution space and enabling the optimizer to balance between $\mathbf{F}_\mathrm{ramp}$ 
and $\mathbf{F}_\mathrm{inj}$. Toward the final stage of the search, $\lambda$ rapidly increases over the course 
of several tens of iterations, reflecting a transition into the convergence phase where the full set of constraints 
is actively enforced.

\begin{figure}[htbp] 
    \centering 
    \includegraphics[width=0.99\linewidth]{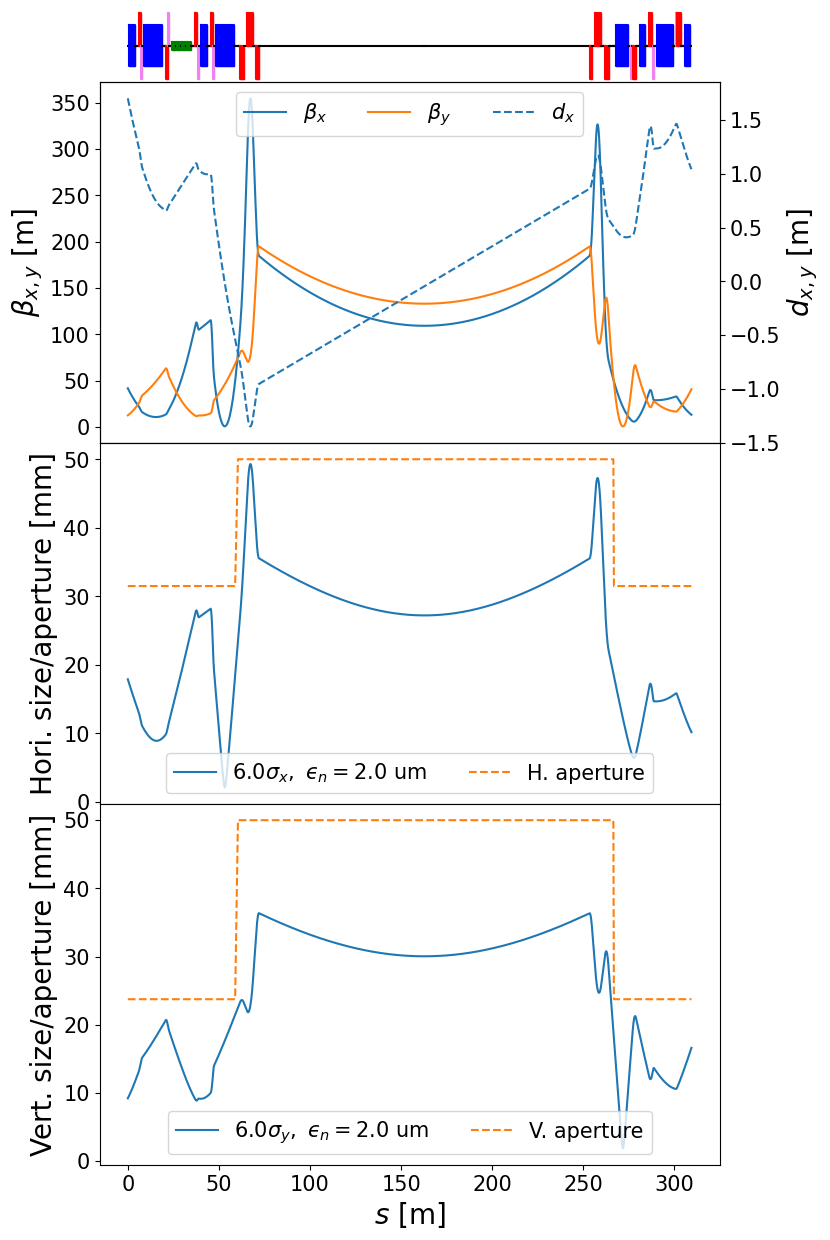} 
    \caption{Injection beam optics ramped down from top-energy configuration in Fig.~\ref{fig:withRC_top}. 
    Top: $\beta$-functions and horizontal dispersion; Middle: horizontal beam size and aperture limit; 
    Bottom: vertical beam size and aperture limit.}
    \label{fig:FitLEC_withRC} 
\end{figure}

By setting $\lambda = 0$, another local optimizer is applied to each solution obtained during the adaptive
iteration. This yields the solution set
\begin{equation}
\mathcal{S} = \{ \mathbf{k} \mid \mathbf{F}_\mathrm{ramp}(\mathbf{k})=\mathbf{0} \}
\end{equation}
which satisfies all ramping constraints across the process. The top-energy optics shown in Fig.\ref{fig:withRC_top} 
and the final matched injection optics shown in Fig.\ref{fig:FitLEC_withRC} are both elements of this set. 
Since our adaptive search algorithm constrains the quadrupole strengths within a small neighborhood during 
each update, the resulting solutions remain close in parameter space. Therefore, a continuous ramping path 
connecting the injection and top-energy configurations can be constructed within $\mathcal{S}$.

\begin{algorithm}[H]
\caption{Constructing a Smooth Ramping Path via Dijkstra and Local Refinement}
\label{alg:ramping_path}
\begin{algorithmic}[1]
\Require Candidate solutions $\mathcal{S}$
\Require Distance threshold $\Delta = 0.1~\mathrm{m}^{-2}$ 
\State Build a graph $G$ with nodes $\mathbf{k}_i\in \mathcal{S}$, 
connect edge if $\|\mathbf{k}_i - \mathbf{k}_j\| < \Delta$
\State Use Dijkstra’s algorithm to find shortest path from top-energy to injection optics
\State Let $\{\mathbf{k}_{s_1}, \ldots, \mathbf{k}_{s_M}\}$ be the selected path
\State Construct cubic spline interpolator over $\{\mathbf{k}_{s_j}\}$ to get $\mathbf{k}(E)$
\For{each energy step $E \in [E_{\text{inj}}, E_{\text{top}}]$}
    \State Use interpolated $\mathbf{k}(E)$ as initial guess
    \State Apply Levenberg-Marquardt optimizer to minimize $\|\mathbf{F}_\text{ramp}(\mathbf{k}(E))\|$
    \State Save refined $\mathbf{k}(E)$
\EndFor
\State \textbf{Output:} Smooth and feasible quadrupole ramp $\{\mathbf{k}(E)\}$
\end{algorithmic}
\end{algorithm}

\begin{figure}[htbp] 
    \centering 
    \includegraphics[width=0.99\linewidth]{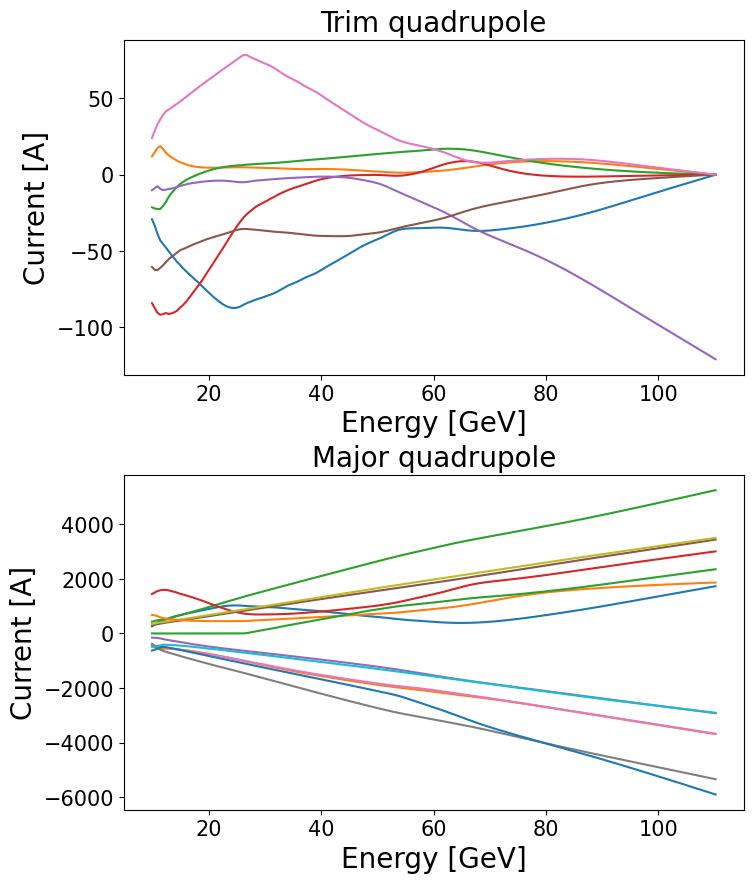} 
    \caption{Quadrupole currents during the energy ramping from injection configuration in Fig.~\ref{fig:FitLEC_withRC} 
    to top energy configuration in Fig.~\ref{fig:withRC_top}, 
    top: the current evolution of the trim 
    quadrupoles, bottom: current of the main quadrupoles. 
    Each curve corresponds to an individual quadrupole in the IR2 region.}
    \label{fig:withRC_ramping} 
\end{figure}

Algorithm~\ref{alg:ramping_path} outlines the procedure for constructing a smooth ramping path from the solution 
set $\mathcal{S}$. A graph is built by connecting nearby points in $\mathcal{S}$, and Dijkstra’s algorithm \cite{dijkstra2022note} is 
applied to identify the shortest path from the top-energy optics to the injection optics. 
This path defines a sequence of intermediate optics configurations, which serve as anchor points for 
cubic spline interpolation. The interpolated quadrupole strengths are then used as initial guesses at each 
discrete energy step.  This ensures that the resulting ramping path remains both smooth and physically feasible 
across the entire energy range. The corresponding quadrupole current profiles during the ramping process are shown 
in Fig.~\ref{fig:withRC_ramping}, illustrating the continuous evolution of both trim and main quadrupoles.

\section{Summary}\label{sec:summary}
This study presents a complete lattice design and optics ramping framework for the IR2 region in the HSR of 
the EIC. The key physical motivation is to provide a long drift section for co-propagating electrons and hadrons, 
which is essential for effective electron cooling. A systematic permutation of RHIC magnet cryomodules is 
compared to maximize the usable drift length.

We show that the optics design problem is fundamentally under-determined: the number of quadrupole knobs exceeds 
the number of matching constraints. The separately matched injection and top-energy optics usually belong to 
disconnected branches in the parameter space. 
This poses a significant challenge when constructing a continuous and hardware-feasible ramping path.

When top energy optics is not heavily constrained, we incorporate a penalty term at intermediate energies to ensure 
continuity and feasibility between injection and top-energy optics. When constraints of top-energy cooling are active, 
we introduce an adaptive optimization scheme that tunes a weighting parameter to gradually enforce 
low-energy requirements while remaining within hardware limits. A shortest-path search in the quadrupole 
parameter space is applied to extract a smooth configuration sequence, and local optimizers are used to 
finalize the ramping path.

The final result satisfies all optics and hardware constraints, while preserving smooth evolution of 
quadrupole strengths. These two methods are general and applicable to other ramping problems where the solution space 
is under-constrained and the top-energy solution lies far from the low-energy configuration in variable space.

\begin{acknowledgments}
The author would like to thank J. S. Berg, G. Robert-Demolaize, and V. Ptitsyn for insightful discussions; 
J. S. Berg, C. Liu, and A. Fedotov for providing key constraints that define the matching process; 
and S. G. Peggs, C. Liu and Y. Li for carefully reviewing the manuscript and offering valuable suggestions.

This work was supported by Brookhaven Science Associates, LLC under Contract No. DE-SC0012704 with the U.S. Department of Energy, and by a Department of Energy Early Career Award.
\end{acknowledgments}

\appendix

\section{Lattice Layout and Geometry Requirements}\label{app:layout}

The placement of the Siberian Snake in IR2 strongly constrains the magnet layout. The Snake and its adjacent quadrupoles occupy about $20~\mathrm{m}$. To maintain $180^\circ$ separation from the IR8 Snake, IR2 must reserve alignment slots for both colliding and non-colliding IR8 configurations~\cite{berg:ipac2023-mopl156, liu:ipac2022-wepopt034}. This reduces the maximum available drift to approximately $160~\mathrm{m}$. Therefore, this study considers only the non-colliding IR8 configuration, consistent with the current EIC baseline. A future detector at IR8 would likely require reconfiguring the IR2 layout.

The HSR-IR2 layout must provide sufficient space for low-energy electron cooling. In this scheme, hadrons co-propagate with electrons at the same average velocity, and Coulomb interactions reduce the hadron emittance~\cite{PhysRevLett.96.044801}. The cooling rate scales with the effective overlap length between the two beams. To satisfy this requirement, a systematic layout search is performed over permutations of dipoles, quadrupoles, the Siberian Snake, and drift elements.

The IR2 region includes eight dipole magnets labeled by the RHIC naming convention: D9, D8, D6, D6, D5O, D5I, D8, 
and D9. In each candidate layout, two dipoles (D9 and D8) are placed before the Siberian Snake and the 
remaining six after it. Accounting for the two identical D6 magnets, the number of unique dipole arrangements 
is $2! \times \frac{6!}{2} = 720$. Seven insertion points between dipoles are available for placing quadrupoles, 
the Snake, and the long drift. There are two possibilities to place the long drift: 
\begin{itemize} 
\item (Q, Snake, Long, Q, Q, Q, Q) 
\item (Q, Snake, Q, Long, Q, Q, Q) 
\end{itemize} 
Here Q denotes a standard RHIC CQS or CQT cryomodule, which contains a standard quadrupole.

The Siberian snake occupies a reserved length of $12.5~\mathrm{m}$. One quadrupole is placed on each side of the 
snake to control the horizontal and vertical beam sizes. The long drift section includes two large-aperture triplets 
at each end. To fully control the six transverse optical parameters at the center of the long drift—namely 
$\beta_x$, $\alpha_x$, $\beta_y$, $\alpha_y$, $\eta_x$, and $\eta_x'$—a minimum of six quadrupoles is required 
on each side of the drift. 
Both candidate layouts satisfy this: one places six quadrupoles left of the long drift and seven on the right; 
the other reverses this arrangement.

Although permutations of quadrupole types could yield more layouts, they are not considered here, as they cause only minor geometric variations.

A short adjustable drift is added to ensure geometric closure. It can be placed in eight locations: before the first 
or after the last dipole (2 options), or adjacent to one of the quadrupoles or the Snake (6 options). 
The total number of configurations is:
\begin{equation}
    720 \times 2 \times 8 = 11{,}520
\end{equation}

The lengths of the short and long drifts are adjusted to meet geometric boundary conditions at YO1 and YI2, the upstream and downstream interfaces to the arc. These conditions, based on the MAD-X \texttt{SURVEY} convention, are listed in Table~\ref{tab:GeometryBoundaryConditions}.

\begin{table}[htbp]
\centering
\caption{Geometry boundary conditions at YO1 and YI2, which mark the upstream and downstream interfaces to 
the arc, respectively. Although the hadron beam circulates counter-clockwise, the lattice is designed in the 
clockwise direction. The symbols $x$, $z$, and $\theta$ follow the MAD-X \texttt{SURVEY} convention~\cite{MADXUserGuide}.}
\label{tab:GeometryBoundaryConditions}
\begin{tabular}{lcc}
\hline
\textbf{Parameter} & \textbf{YO1} & \textbf{YI2} \\
\hline
$x~[\mathrm{m}]$         & $427.705516$  & $159.486753$ \\
$z~[\mathrm{m}]$         & $431.526825$  & $585.348499$ \\
$\theta~[\mathrm{rad}]$  & $-0.951763070$ & $-1.142632032$ \\
\hline
\end{tabular}
\end{table}

All $11,520$ configurations are evaluated. A valid layout requires positive drift lengths. Among feasible solutions, 
$777$ provide a long drift exceeding $170~\mathrm{m}$. The configuration with the longest usable drift is shown 
in Fig.~\ref{fig:HSR-IR2-layout}.

\section{Beam Optics Constraints}\label{app:constraints}
The Twiss functions must match the upstream and downstream arc optics throughout ramping. 
The required boundary conditions are listed in Table~\ref{tab:CombinedOptics}.
At the center of the long drift, the beam optics are optimized for electron cooling. At injection, the beta functions should lie within $100 \sim 200~\mathrm{m}$, with near-zero dispersion, consistent with the standalone LEC 
design derived from LEReC experience at RHIC~\cite{fedotov2020experimental}.
At top energy, two scenarios are considered. Without high-energy cooling, the optics at the drift center 
remain unconstrained. If the Ring Cooler is implemented, the optics must meet stricter requirements, 
including large vertical beta function $\beta_y^*>800~\mathrm{m}$ and large horizontal dispersion $\eta_x^*\sim 2~\mathrm{m}$. 
These target values are also summarized in Table~\ref{tab:CombinedOptics}.

\begin{table}[htbp]
\centering
\caption{Twiss parameters at the upstream (YO1), downstream (YI2), and center of the long drift. The values at YO1 
and YI2 match arc optics across all energies; the values at the center of the long drift indicate optimal ranges
for electron cooling at injection energy (LEC Center), or at top energy (RC Center).}
\label{tab:CombinedOptics}
\begin{tabular}{lcccc}
\hline
\textbf{Parameter} & \textbf{YO1} & \textbf{YI2} & \textbf{LEC Center} & \textbf{RC Center}\\
\hline
$\beta_x$ [m]   & $41.66$   & $13.18$   & $[100, 200]$ & $200$\\
$\alpha_x$      & $1.97$    & $0.74$    & $0$ & $0$\\
$\eta_x$ [m]    & $1.70$    & $1.04$    & $[-1, 1]$ & $\sim2$\\
$\eta_x'$       & $-0.0825$ & $-0.044$  & $[-0.02, 0.02]$ & $0$\\
$\beta_y$ [m]   & $12.53$   & $40.70$   & $[100, 200]$ & $>800$\\
$\alpha_y$      & $-0.71$   & $-1.97$   & $0$ & $0$\\
\hline
\end{tabular}
\end{table}

The physical aperture is limited by the RHIC vacuum chambers, upgraded with copper-coated beam screens to 
reduce resistive wall heating and suppress electron cloud effects~\cite{verdu-andres:ipac2021-tupab260}. 
A $6\sigma$ beam envelope is used as the design limit. This is challenging at injection due to large beta functions 
and the large emittance of low-energy beams. 
Large-aperture triplets are placed at both ends of the long drift to reduce beam size and control beta functions.
The gold ion beam defines the most stringent aperture constraints and is used as the reference case. 
Table~\ref{tab:ApertureConstraints} summarizes the relevant beam parameters and aperture limits in this paper.

\begin{table}[htbp]
\centering
\caption{Aperture constraints at injection. The $6\sigma$ beam envelope must remain within the physical aperture.}
\label{tab:ApertureConstraints}
\begin{tabular}{lc}
\hline
\textbf{Parameter} & \textbf{Value} \\
\hline
Species & Golden ion beam \\
Beam energy, $E$                        & $10~\mathrm{GeV}$ \\
Normalized emittance, $\epsilon_n$     & $2.0~\mathrm{\upmu m}$ \\
Relative momentum spread, $\sigma_\delta$ & $6 \times 10^{-4}$ \\
Aperture at triplets                    & $6\sigma_x, 6\sigma_y \leq 50~\mathrm{mm}$ \\
\multirow{2}*{Aperture at other magnets}            & $6\sigma_x \leq 31.5~\mathrm{mm}$\\
& $6\sigma_y \leq 23.75~\mathrm{mm}$ \\
\hline
\end{tabular}
\end{table}

As shown in Fig.~\ref{fig:HSR-IR2-layout}, the IR2 lattice uses two types of quadrupoles: standard RHIC quadrupoles 
(“main”) and individually powered trim quadrupoles. Each has a distinct transfer function due to differences 
in magnetic design and power supply configuration \cite{RHICConfigManual}.
Table~\ref{tab:QuadStrengthLimits} lists the quadrupole current following the RHIC operational practice and 
the corresponding normalized focusing strengths at top energy ($110~\mathrm{GeV}$ gold beam).
\begin{table}[htbp]
\centering
\caption{Maximum quadrupole strength at top energy. The normalized focusing strength is defined as
$k=\frac{1}{B\rho}\frac{\partial B_y}{\partial x}$.}
\label{tab:QuadStrengthLimits}
\begin{tabular}{lcc}
\hline
\textbf{Quadrupole Type} & \textbf{Max Current [A]} & \textbf{$k~[\mathrm{m}^{-2}]$} \\
\hline
main quadrupole & 6000 & 0.098 \\
Trim quadrupole  & 150  & 0.046 \\
\hline
\end{tabular}
\end{table}

\bibliography{ref}

\end{document}